\documentclass{PoS}

\title{Unveiling the submerged part of the iceberg:\\
 radio-loud narrow-line Seyfert 1s with SKA}

\ShortTitle{RLNLS1s with SKA}

\author{\speaker{M. Berton} $^{1}$, L. Foschini$^{2}$, A. Caccianiga$^{2}$, J.~L. Richards$^3$, S. Ciroi$^1$, E. Congiu$^1$, V. Cracco$^1$, G. La Mura$^1$, L. Marafatto$^4$, P. Rafanelli$^1$\\
        $^{1}$ Dipartimento di Fisica e Astronomia "G. Galilei", Universit\`a di Padova, Vicolo dell'osservatorio 3, 35122, Padova, Italy\\
	$^{2}$ INAF - Osservatorio Astronomico di Brera, Via E. Bianchi 46, 23807, Merate (LC), Italy;\\
	$^{3}$ Department of Physics and AStronomy, Purdue University, 525 Northwestern Avenue, West Lafayette, IN 47907, USA; \\
	$^4$ INAF - Osservatorio Astronomico di Padova, Vicolo dell'osservatorio 5, 35122, Padova, Italy.\
        E-mail: \email{marco.berton.1@studenti.unipd.it}}

\abstract{Narrow-line Seyfert 1 galaxies (NLS1) are active galactic nuclei (AGN) known to have small masses of the central black hole and high accretion rates. NLS1s are generally radio-quiet, but a small part of them (about 7\%) are radio-loud. The recent discovery of powerful relativistic jets in radio-loud NLS1s (RLNLS1s), emitting at high-energy $\gamma$-rays, opened intriguing questions. The observed luminosity of the jet is generally weak, smaller than blazars, although when rescaled for the mass of the central black hole, it becomes of the same order of magnitude of the latter. The weak luminosity, and hence observed flux, resulted in a small number of known RLNLS1. From a recent survey of RLNLS1s, it was found that only 8 out of 42 sources had radio flux density at 1.4 GHz greater than 100 mJy, while 21 out of 42 had flux density smaller than 10 mJy. In addition, given the strong variability at all wavelengths, with present-day facilities RLNLS1s can often only be detected during high activity periods. The Square Kilometer Array (SKA), with its superior sensitivity, will break this limit, allowing us to unveil a relatively unknown population of jetted AGN. We present the results of a study aimed at evaluating the scenario that could emerge after the advent of SKA.
	}

\FullConference{EXTRA-RADSUR2015 (*)\\
		20--23 October 2015\\
		Bologna, Italy

                \bigskip
                \hrule
                \bigskip

                \textnormal{(*) This conference has been organized
                  with the support of the Ministry of Foreign Affairs
                  and International Cooperation, Directorate General
                  for the Country Promotion (Bilateral Grant Agreement
                  ZA14GR02 - Mapping the Universe on the Pathway to
                  SKA)}
}

\begin{document}

\section{Introduction}

Narrow-line Seyfert 1 galaxies (NLS1s) are a class of active galactic nuclei (AGN) that since its classification \cite{Osterbrock85} has always presented new, intriguing challenges for AGN physics and modelling. They are classified on the basis of the full width at half maximum (FWHM) of the H$\beta$ line, FWHM(H$\beta$) $<$ 2000 km s$^{-1}$, and of the ratio between [O III]/H$\beta$ $<$ 3. The width of permitted lines is comparable to that of type 2 AGN, but the presence of strong Fe II multiplets in the spectrum reveals that the BLR is actually visible, as in type 1 AGN. The lines narrowness is therefore attributed to a low rotational velocity around a relatively low mass black hole (10$^6$-10$^8$ M$_\odot$, \cite{Mathur00}). Their bolometric luminosity is similar to that of broad-line Seyfert 1, and this translates in a very high Eddington ratio, possibly a sign of a very high accretion rate \cite{Boroson92}. For all these reasons, NLS1s are sometimes considered a young and fastly growing phase in type 1 AGN evolution \cite{Grupe00, Mathur00}.\par
In radio NLS1s do not usually exhibit any intense emission. Indeed 93\% of them are radio-quiet, while the remaining 7\% is radio-loud \cite{Komossa06}. Some of these radio-loud NLS1s (RLNLS1s) show several blazar-like properties, such as a high brightness temperature and a flat-radio spectrum \cite{Yuan08}, and in recent years the Fermi Satellite discovered $\gamma$-ray emission coming from 10 of them, indicating the presence of a relativistic beamed jet \cite{Abdo09a, Foschini15, Dammando15, Yao15, Liao15}. Misaligned jets were also found in a few steep radio-spectrum NLS1s (S-NLS1s, e.g. \cite{Caccianiga14, Richards15}), and these steep-spectrum sources were pointed out as part of the beamed sources parent population \cite{Berton15}. \par
One of the biggest issues in the study of radio-emitting NLS1s is the small number of sources available. As shown in the following section, we found only 149 radio-emitting NLS1s up to z = 0.3. The aim of this work is to investigate how next generation instruments, and in particular the Square Kilometer Array (SKA), will affect the search for this elusive class of AGN. 

\section{Samples selection}
We decided to use two samples of relatively low $z$ NLS1s. We extracted from SDSS DR7 all the NLS1s at $z < 0.3$, using as classification criteria the FWHM(H$\beta) <$ 2000 km s$^{-1}$ and the ratio [O III]/H$\beta < 3$. Then we searched for a FIRST radio-source within a radius of 5 arcsec \cite{Becker95}. Finally we calculated their radio-loudness, defined as RL = F$_{B-band}$/F$_{5 \; GHz}$ \cite{Kellermann89}. To derive the B-band magnitude, we convolved the optical spectra with a B-band filter template, and then estimated the integrated flux. The 5 GHz flux instead was obtained from the 1.4 GHz flux of FIRST assuming a spectral index $\alpha = 0.5$ (F$_\nu \propto \nu^{-\alpha}$ \cite{Yuan08}).
Sources with RL $>$ 10 were defined as radio-loud, otherwise they were considered radio-quiet. In this way we obtained 117 RQNLS1s, and 32 RLNLS1s. We tested the samples by means of V/V$_{max}$ test \cite{Avni80} to investigate their completeness. The radio-quiet sample has $\langle$V/V$_{max}\rangle$ = 0.54$\pm$0.03, while the radio-loud has $\langle$V/V$_{max}\rangle$ = 0.45$\pm$0.05. Both the samples are then in agreement with the uniform distribution of sources, and they can be considered statistically complete. \par
\section{Results}
\begin{figure*}[t!]
\centering
\includegraphics[width=7cm]{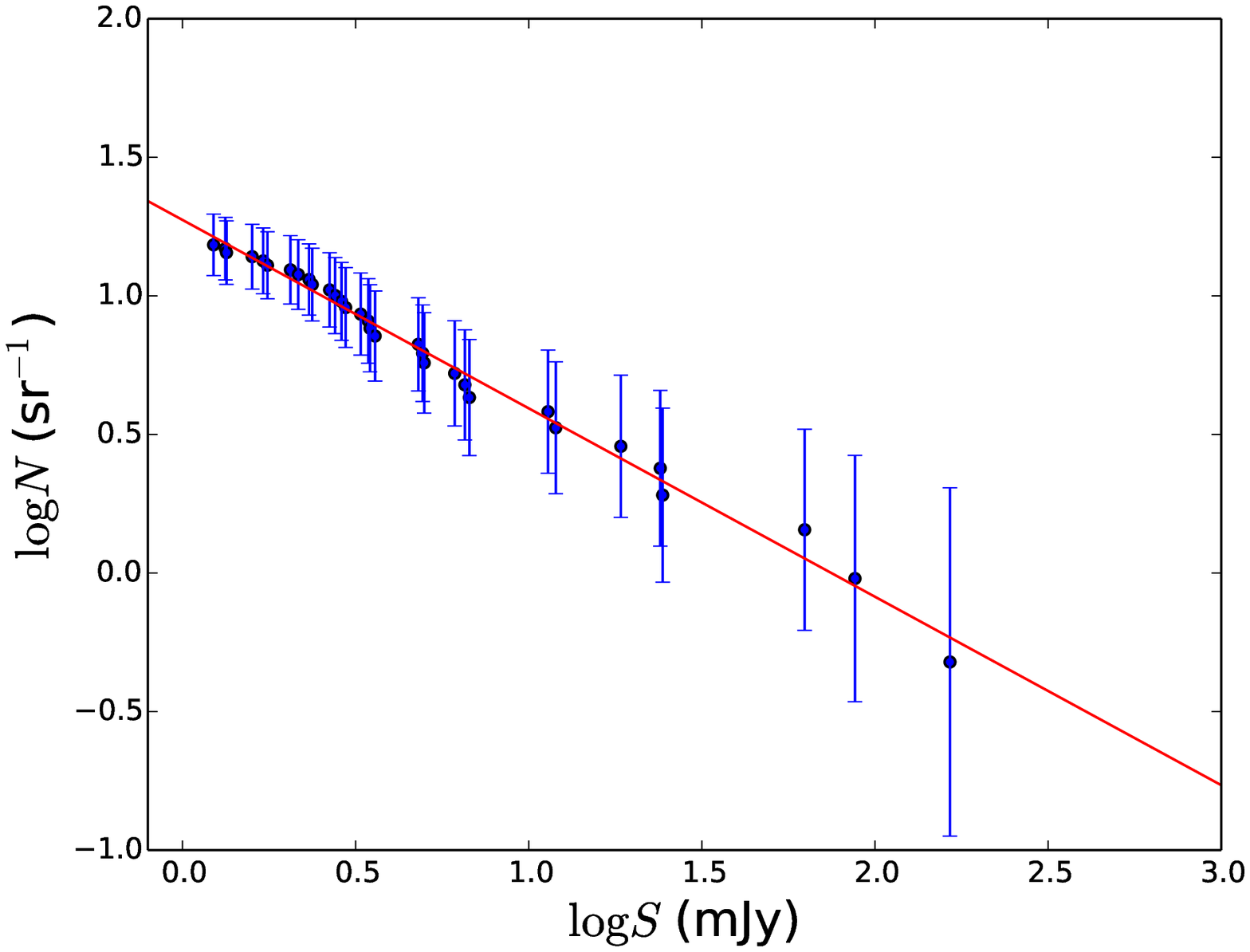} 
\includegraphics[width=7cm]{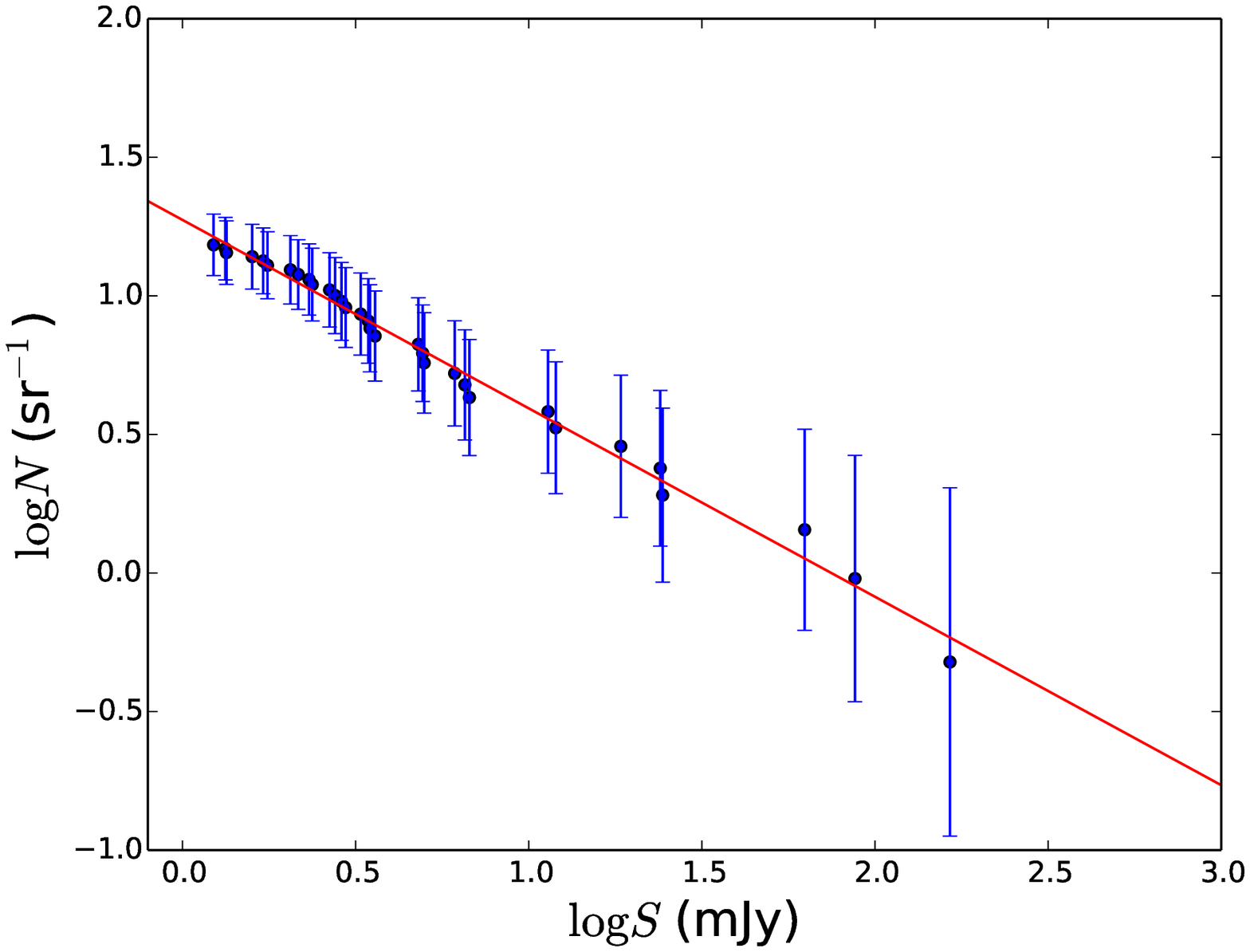}
\caption{$\log$N-$\log$S test for radio-loud (left panel) and radio-quiet (right panel) NLS1s. The errors are poissonian. The red solid line indicates the linear best-fit.}
\label{logNlogS}
\end{figure*}
A useful tool to investigate the space density of sources is the $\log$N-$\log$S test, which does not require the knowledge of the distance of each source. The test assumes that all the sources are above a minimum luminosity which always corresponds to an observable flux in the explored volume, an assumption that appears reasonable for samples limited to relatively low redshifts. The test evaluates the number of sources above each flux density $S$, so it studies the cumulative distribution of sources as a function of the flux. The number of sources brighter than some flux $S$ should then be proportional to $S^{-3/2}$, for constant density in a Euclidean space. This relationship is known as the $\log$N-$\log$S test, which was developed to test the evolution of radio sources, so it is particularly well suited for our aim. \par
The sources counts are evaluated by considering the theoretical observing limit of SKA in declination ($\sim$60$^\circ$ N), and the SDSS DR7 sky coverage ($\sim$1/6 of the whole sky). The results are shown in Fig.~\ref{logNlogS}. The errors are Poissonian. The distribution of RQNLS1s is the closer one to the Euclidean distribution (slope $\beta = 1.64\pm0.01$), while RLNLS1s have a flatter distribution ($\beta = 0.68\pm0.01$). The SKA-1 sensitivity\footnote{SKA-TEL-SKO-0000229 - Report and Options for Re-Baselining of SKA-1} at 1.4 GHz will be 0.082 mJy hr$^{-1/2}$. In our calculations we assume that the resulting slope is remaining the same at lower fluxes. This can be done in absence of evolutionary effects, that we do not expect to be particularly strong in our relatively low redshift domain. This simple extrapolation yields very large numbers: SKA might be able to identify $\sim$1300 RLNLS1 and $\sim$50000 RQNLS1s up to $z=0.3$. \par
This important result, nevertheless, is only an upper limit, since there are a few caveats that must be kept in mind. In particular, regarding RLNLS1s, a decrease of one order of magnitude in radio flux must correspond to a similar decrease also in optical flux, otherwise the source would not be radio-loud anymore. It is then possible that SKA will not detect many more RLNLS1s, at least up to z = 0.3, although it will clearly allow a much deeper study of all currently known RLNLS1s. Conversely, at higher $z$, the number of new RLNLS1s can be largely increased, since there are many radio-loud sources with a flux density $\sim$1 mJy that at larger distances are not visible for present day observatories. \par
For RQNLS1s there are also a couple of caveats. First, it is unlikely that so many NLS1s exist up to $z = 0.3$, but this only means that SKA will be able to detect all the existing NLS1s. But whilst SKA will detect many more sources, it will be difficult to identify them as NLS1s without a deep optical observation. The radio luminosity is proportional to the [O III] line luminosity \cite{Debruyn78}, hence a low radio flux corresponds to an equally low [O III] flux. This makes the classification more and more difficult for weak sources. \par
Anyway, even with this optical limit, the knowledge increase on the nature of RQNLS1s will be very large. In particular, a deep radio investigation will provide helpful information on the dichotomy between radio-quiet and radio-loud sources. In turn it will also help us to unveil whether the origin of radio emission in RQNLS1s is due to the high star formation rate typical of NLS1s \cite{Sani10}, or to some sort of weak activity that might be ongoing, such as an aborted or faint jet \cite{Ghisellini04, Doi13}. Nonetheless, it will also allow to study in detail the radio emission from RLNLS1s, investigating at very high resolution the morphology of the jets, and finally clarifying how strong the incidence of the starburst component is in these sources \cite{Caccianiga15}. Last but not least, SKA will provide enough statistic to perform a detailed investigation on RLNLS1s by means of the radio luminosity function, allowing us to study their evolution through cosmic time.


\begin{thebibliography}{99}
\bibitem{Abdo09a} Abdo, A.A., et al.\ 2009, ApJ, 699, 976
\bibitem{Avni80} Avni, Y., \& Bahcall, J.~N.\ 1980, ApJ, 235, 694 
\bibitem{Becker95} Becker, R.~H., et al.\ 1995, ApJ, 450, 559 
\bibitem{Berton15} Berton, M., et al., 2015\ A\&A, 578, A28
\bibitem{Boroson92} Boroson, T.A., \& Green, R.F.\ 1992, ApJS, 80, 109
\bibitem{Caccianiga14} Caccianiga, A., et al. \ 2014, MNRAS, 441, 172
\bibitem{Caccianiga15} Caccianiga, A., et al. \ 2015, MNRAS, 451, 1795
\bibitem{Dammando15} D'Ammando, F., et al.\ 2015, MNRAS, 452, 520
\bibitem{Debruyn78} de Bruyn, A.~G., \& Wilson, A.~S.\ 1978, A\&A, 64, 433
\bibitem{Doi13} Doi, A., et al.\ 2013, ApJ, 765, 69
\bibitem{Foschini15} Foschini, L., et al.\ 2015, A\&A, 575, A13
\bibitem{Ghisellini04} Ghisellini, G., et al.\ 2004, A\&A, 413, 535
\bibitem{Grupe00} Grupe, D.\ 2000, New. Astron. Rev., 44, 455
\bibitem{Kellermann89} Kellermann, K.~I., et al.\ 1989, AJ, 98, 1195 
\bibitem{Komossa06} Komossa, S., et al.\ 2006, AJ, 132, 531
\bibitem{Liao15} Liao, N.-H., et al.\ 2015, arXiv:1510.05584 
\bibitem{Mathur00} Mathur, S.\ 2000, MNRAS, 314, L17
\bibitem{Osterbrock85} Osterbrock, D.E., \& Pogge, R.W.\ 1985, ApJ, 297, 166
\bibitem{Richards15} Richards, J.L., \& Lister, M.L.\ 2015, ApJ, 800, L8
\bibitem{Sani10} Sani, E., et al.\ 2010, MNRAS, 403, 1246 
\bibitem{Yao15} Yao, S., et al.\ 2015, MNRAS, 454, 16
\bibitem{Yuan08} Yuan, W., et al.\ 2008, ApJ, 658, 801

\end{thebibliography}
\end{document}